\documentclass[twocolumn,prl,amsmath,amssymb,showpacs,superscriptaddress,floatfix]{revtex4}
\usepackage{graphicx}

\usepackage{color}
\usepackage{amsmath}
\usepackage{bm}

\renewcommand{\vec}[1]{\mathbf{#1}}

\begin{document}

\title{High-temperature surface superconductivity in rhombohedral graphite}

\author{N.~B.~Kopnin}

\affiliation{Low Temperature Laboratory, Aalto University, P.O. Box 15100, FI-00076 AALTO, Finland}

\affiliation{ L.~D.~Landau Institute for
Theoretical Physics, 117940 Moscow, Russia}

\author{M. Ij\"as}
\author{A. Harju}
\affiliation{COMP Centre of Excellence and Helsinki Institute of Physics, Department of Applied Physics, Aalto University, P.O. Box 14100, FI-00076 AALTO, Finland}

\author{T.~T.~Heikkil\"a}

\affiliation{Low Temperature Laboratory, Aalto University, P.O. Box 15100, FI-00076 AALTO, Finland}

\date{\today}

\begin{abstract}
Surface superconductivity in rhombohedral graphite is a robust phenomenon which can exist even when higher order hoppings between the layers lift the topological protection of the surface flat band and introduce a quadratic dispersion of electrons with a heavy effective mass. We show that for weak pairing interaction, the flat band character of the surface superconductivity transforms into a BCS-like relation with high critical temperature characterized by a higher coupling constant due to a much larger density of states than in the bulk. Our results offer an explanation for the recent findings of graphite superconductivity with an unusually high transition temperature.
\end{abstract}


\maketitle

A low critical temperature of conventional superconductors results from a constant density of states (DOS) due to a linear-in-momentum electronic spectrum near the Fermi energy. With a higher-order dispersion, the relation between the critical temperature and the coupling constant becomes stronger, boosting the superconductivity. The extreme case would be a completely dispersionless energy spectrum, a flat band, which has been predicted in
many condensed matter systems, see e.g.
Refs.~\onlinecite{Khodel1990,NewClass,Shaginyan2010,Gulacsi2010}. In
some cases the flat bands are protected by topology in momentum
space; they emerge in gapless topological
matter\cite{HeikkilaKopninVolovik10,Ryu2002,SchnyderRyu2010,GuineaCNPeres06,HeikkilaVolovik10-1,MakShanHeinz2010,Dora2011,KopninSalomaa1991,Volovik2011}.
A singular DOS associated with the
dispersionless spectrum was recently shown\cite{KopninHeikkilaVolovik2011} to essentially enhance the transition
temperature opening a new route to room-temperature
superconductivity.

The problem is to find the metal with such a higher-order dispersion around the Fermi sea. Refs.~\onlinecite{HeikkilaVolovik10-1} and \onlinecite{KopninHeikkilaVolovik2011} have shown that within the nearest-neighbour approximation, rhombohedral graphite (RHG) has topologically protected surface states with a flat band at the Fermi energy, and these surface states support high-temperature superconductivity where the superconducting order parameter is concentrated around the surfaces. A flat band forms out of a low dispersive band that appears on the surface of a multilayered rhombohedral graphene structure with a large number of layers. The corresponding critical temperature depends linearly on the pairing interaction strength and can be thus
considerably higher than the usual exponentially small critical
temperature in the bulk. Flat-band superconductors can carry quite high surface supercurrent with the critical value proportional to the critical temperature \cite{Kopnin2011}.

Experimental evidence of a high-temperature superconductivity in graphite in the form of a small Meissner effect and of a sharp drop in resistance appeared in the literature during past years \cite{Kopelevich01,Esquinazi08}.  Recently, these findings have been ratified by observations of zero resistance in graphitic samples up to 175 K \cite{Ballestar12} and indications of even room-temperature superconductivity in specially prepared graphite samples \cite{Esquinazi12}. The enhanced superconducting density has been also reported on twin boundaries in Ba(Fe$_{1-x}$Co$_x$)$_2$As$_2$ \cite{Moler2010}. In this Letter, we argue that the high-temperature superconductivity in graphite can be related to surface superconductivity that may form either on the outer surfaces of the sample or on twin boundaries of or on grain boundaries between inclusions of RHG. We demonstrate that the surface superconductivity is a robust phenomenon which survives even when the topological protection of the flat band itself is lifted. In particular, the next-nearest neighbour hoppings in RHG can break the exact topological protection and, therefore, the flat-band mechanism of superconductivity could be destroyed. We study these higher-order interactions \cite{KopninHeikkila12} and show that, though breaking the flat-band scenario at sufficiently low values of the coupling energy, they provide another mechanism of surface superconductivity which is of the BCS type but still has a much larger coupling constant than the usual superconductivity in bulk, thus favoring high-temperature superconductivity. The enhanced coupling constant comes from a high DOS associated with a heavy effective mass of surface quasiparticles emerging on the background of the pre-existing flat band. Our results help to identify the regime of parameters where extremely high-temperature surface superconductivity may be found.

\begin{figure}[t]
\includegraphics[width=0.6\linewidth]{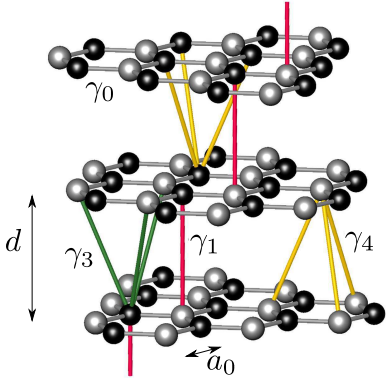}
\caption{Rhombohedral graphite. The black and gray atoms correspond to A and B sites, respectively. } 
\label{fig:rhggraphite}
\end{figure}

\paragraph{Electron dispersion in RHG.} 

The RHG lattice and the tight-binding couplings are depicted in Fig.~\ref{fig:rhggraphite}. We label the layers (starting from the bottom) by $n$, the atoms A in layer $n$ are on top of atoms B in layer $n-1$, and the vector between the layers is $\vec{d}$. 
In our estimates, we use the tight-binding parameters denoted in Fig.~\ref{fig:rhggraphite}. They satisfy $\gamma_0 \gg \gamma_1 \sim \gamma_3 \gg \gamma_4$ \cite{reviewCastroNeto09}. In the tight-binding numerics below, we use the values $\gamma_0=2.58$ eV, $\gamma_1=0.34$ eV, $\gamma_3=0.17$ eV, and $\gamma_4=0.04$ eV, which give the best fit to the density functional theory (DFT) calculation of the surface state dispersion (see Appendix).

\begin{figure}[t]
\includegraphics[width=0.8\linewidth]{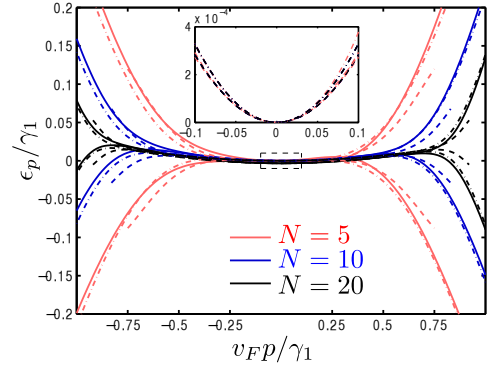}
\caption{Cuts of the 3d spectrum for $p<p_{\rm FB}\equiv \gamma_1/v_F$ along the K-$\Gamma$ (K-M) on negative (positive) x-axis; $N=5$ (red), $N=10$ (blue), and $N=20$ (black) graphene layers. The DFT calculations are shown in full, tight-binding in dash-dotted and  Eq.~\eqref{eq:normaldispersion}  in dashed lines (the latter plotted up to their regimes of applicability). The inset is a zoom-up of the low-energy region with tight-binding (dash-dotted) and analytical (dashed) curves. The deviations between the dashed and other lines show up when $\xi_p$ becomes dominant in Eq.~\eqref{eq:normaldispersion}, and are partially due $\gamma_3$ neglected there.} 
\label{fig:2dnormalspectrum}
\end{figure}

The RHG is a multilayered graphene structure. The conical spectrum near the Dirac points ${\bf K}$ and ${\bf K}^\prime$ of the Brillouin zone of a single-layer graphene (for details, see review \cite{reviewCastroNeto09} and references therein) is transformed into low-dispersion, low-energy bands (see Fig.~\ref{fig:2dnormalspectrum}) which determine the unique features of this system.
Since we are interested in low energies  we concentrate on the in-plane momenta ${\bf p}=(p_x, \ p_y)$ close to one of these Dirac corners. A standard Fourier series expansion near ${\bf K}$ yields \cite{KopninHeikkila12}
\begin{eqnarray}
H_{{\bf K}}=\sum_{{\bf p}}\sum_{m,n=1}^N \hat \psi_m^\dagger ({\bf p}) \hat H_{mn}({\bf K},{\bf p}) \hat \psi_n  ({\bf p}), \label{H-particle}
\end{eqnarray}
where 
$\hat H_{mn}({\bf K},{\bf p})=\sum_{l=0}^4 \hat H_{mn}^{(l)}({\bf K},{\bf p})$ and
\begin{eqnarray*}
\hat H_{mn}^{(0)}({\bf K},{\bf p})\!\!&=&\!\! v_F (\hat {\bm \sigma}\cdot {\bf p})\delta_{mn} \\
\hat H_{mn}^{(1)}({\bf K},{\bf p})\!\!&=&\!\! -\gamma_1 \left[ e^{-i\frac{\pi}{6}} \hat \sigma_+ \delta_{m,n+1}+  e^{i\frac{\pi}{6}} \hat \sigma_- \delta_{m,n-1}\right]  \\
\hat H_{mn}^{(3)}({\bf K},{\bf p})\!\!&=&\!\! \tilde\gamma_3 v_F \left[e^{-i\frac{\pi}{3}}   \hat \sigma_+ p_+  \delta_{m,n-1}
+e^{i\frac{\pi}{3}}  \hat \sigma_- p_-  \delta_{m,n+1}  \right]\\
\hat H_{mn}^{(4)}({\bf K},{\bf p})\!\!&=&\!\! \tilde\gamma_4  v_F \left[e^{i\frac{\pi}{6}}   p_-  \delta_{m,n-1} + 
e^{-i\frac{\pi}{6}}  p_+ \delta_{m,n+1} \right].
\end{eqnarray*}
Here $\tilde \gamma_3 =\gamma_3/\gamma_0$, $\tilde \gamma_4 =\gamma_4/\gamma_0$,  
$ p_\pm = p_x \pm i p_y =pe^{\pm i\phi}$, and
$
v_F= 3a_0\gamma_0/2\hbar
$.
The Pauli matrices $\hat{\bm \sigma}$ and $2\hat \sigma_\pm =\hat \sigma_x \pm i\hat \sigma_y$ act on pseudo-spinors
$\hat \psi_n = (  \psi_n^{1}\ , \; \psi_n^{2} )^T$, $\hat \psi_n^\dagger = (  \psi_n^{1*}\ , \; \psi_n^{2*} )$,
where $\psi^1_n = \psi_n^A $, $\psi^2_n=e^{i\pi/6} \psi_n^B$.

To construct the associated Bogoliubov--de Gennes (BdG) Hamiltonian for the superconducting state we need also the time-reversed 
``hole'' Hamiltonian for the Dirac point ${\bf K}$. It follows from the particle Hamiltonian in a vicinity of the opposite Dirac point $-{\bf K}$ which is equivalent to ${\bf K}^\prime$.
The wave function $\psi_{{\bf K}}^h $ of a hole excitation near ${\bf K}$ is  $\psi_{{\bf K}}^h = \bar \psi_{-{\bf K}}^*$. One can check that the hole Hamiltonian is $H_{mn}^h({\bf K},{\bf p})=H_{mn}^*(-{\bf K},-{\bf p})$. Therefore,
\begin{equation}
H^h_{{\bf K}} 
 = \sum_{{\bf p}}\sum_{m,n=1}^N \hat {\psi}_m^{(h) \dagger } ({\bf p}) \hat H_{mn}({\bf K},{\bf p}) \hat {\psi}_n^{(h)}  ({\bf p})\ . \label{H-hole}
\end{equation}
In what follows, we denote the electron wave function by $\hat u_n = \hat \psi_n$  and the hole wave function by $\hat v_n =\hat \psi^h_n$. In the next section we study the normal state using Hamiltonian Eq. (\ref{H-particle}) and the relative magnitudes of the coupling constants listed above. The results of the numerical solution are displayed in Fig.~\ref{fig:2dnormalspectrum}, along with the corresponding analytical approximations and DFT calculations.

\paragraph{Low-energy spectrum in the normal state.}

The Schr{\"o}dinger equation takes the form
\begin{equation}
\sum _m \hat H_{nm}({\bf K},{\bf p}) \hat u_m ({\bf p}) =(\epsilon +\mu) \hat u_n({\bf p})\ . \label{Schr-eq1}
\end{equation}
The energy is measured from the chemical potential $\mu$. The energy spectrum in bulk is obtained by ignoring the outermost layers  $n= 1$ and $N$ and using the ansatz 
$
\hat u_n \propto e^{ip_z dn}
$,
where $p_z$ is out-of-plane momentum. 
For zero doping $\mu =0$, the Fermi surface is determined by
$\epsilon(p,p_z,\phi)=0$. If $\gamma_3 = \gamma_4 =0$, the Fermi surface shrinks to a spiral line
$
v_Fp= \gamma_1\ , \; \phi =p_zd+\pi/6$. Projection of this spiral onto the momentum plane $q=0$ determines the area of a flat band for surface states \cite{HeikkilaVolovik10-1} in the limit $N\to \infty$. If only $\gamma_4 =0$ while $\gamma_3\ne 0$, equation $\epsilon =0$  for $\mu=0$ can still be shown to give a Fermi surface in the form of a (corrugated) spiral whose projection determines the surface flat band \cite{KopninHeikkila12}. This is because the interaction comes with matrices $\hat \sigma_x$ and $\hat \sigma_y$  so that the full Hamiltonian obeys the same anti-commutation rule 
$[ \hat \sigma_z ,(\hat H^{(0)} +\hat H^{(1)}+ \hat H^{(3)}) ]_+ =0$
as the initial Hamiltonian $ \hat H^{(0)} +\hat H^{(1)}$.  This preserves the same topological invariant and the same topology of the Fermi surface \cite{HeikkilaVolovik10-1}. Since $\gamma_3$ does not affect the presence of a flat band, we restrict our analytical consideration to the case when only $\gamma_4$ is nonzero while $\gamma_3=0$ for simplicity. However, our numerical analysis is carried out using the full Hamiltonian.

Surface states have complex $p_z=p_z^\prime +i p_z^{\prime\prime}$ and decay into the bulk. For $\gamma_3=0$ Eq. (\ref{Schr-eq1}) in the particle channel,
\begin{eqnarray}
v_F  (\hat {\bm \sigma}\cdot {\bf p}) \hat u_n ({\bf p}) -\gamma_1 \left[ e^{i\frac{\pi}{6}} \hat \sigma_- \hat u_{n+1} + e^{-i\frac{\pi}{6}} \hat \sigma_+ \hat u_{n-1}\right] \nonumber \\
+\tilde\gamma_4  \left[ e^{i\frac{\pi}{6}} v_Fp_- \hat u_{n+1} +  e^{-i\frac{\pi}{6}} v_Fp_+ \hat u_{n-1}\right] =(\epsilon +\mu)\hat u_n \ ,\label{Schr-eq2}
\end{eqnarray}
for low energies has a solution in the form 
\begin{eqnarray}
\hat u_n &=& C e^{i(\phi -\frac{\pi}{6})(n-1-\frac{N}{2})}\nonumber \\
&\times& \left[ \tilde p^{n-1}\,\left(\begin{array}{c} 1 \\ \zeta e^{i\phi} \end{array}\right)A_+ + \tilde p^{N-n}\,\left(\begin{array}{c} \zeta \\ e^{i\phi}\end{array}\right)A_-\right] \quad \label{psi}
\end{eqnarray}
where $\tilde p =p/p_{\rm FB}$, $p_{\rm FB}=\gamma_1/v_F$ and
\begin{equation}
\zeta = \tilde p[(\epsilon +\mu)/\gamma_1 -\tilde \gamma_4 (\tilde p^2+1)]/(\tilde p^2-1)\ . \label{zeta}
\end{equation}
Here the out-of-plane momentum $p_z^\prime d = \phi-\pi/6$ while $e^{\pm p_z^{\prime\prime}d} =\tilde p$.
The overall normalization $C$ is found from $d \sum_{n=1}^N[{\rm Tr}\, \hat u^\dagger_n \hat u_n ]=1$. For large $N$ this gives
$|C |^2=d^{-1}[1-\tilde p^2] $
provided $|A^+|^2+|A^-|^2 =1$.

At the outermost layers, the terms with $\hat \psi_0$ and $\hat \psi_{N+1}$ in Eq. (\ref{Schr-eq2}) disappear.
The components which do not have $\gamma_1$ couple the constants $A^+$ and $A^-$ in Eq. (\ref{psi}) and determine the energy of the surface states
\begin{equation}
\epsilon_p = \mu_p  \pm \xi_p \left(1-\tilde p^2\right)\ , \; \xi_p=\gamma_1 \tilde p^N
\label{eq:normaldispersion}
\end{equation}
for $\xi_p , \epsilon \ll \gamma_1$. Here
\begin{equation}
\mu_p= p^2/2m^* -\mu\ ,\;
m^*=\gamma_1/(4 \tilde \gamma_4 v_F^2) \ . \label{mass}
\end{equation}
The interaction $\gamma_4$ breaks the symmetry between the conduction and valence bands in a way similar to a shift in $\mu$ due to doping. The spectrum $\epsilon_p$ has a quadratic dispersion with the effective mass $m^*$ on a background of a much weaker high-order dispersion $\xi_p$. The latter transforms into a flat band $\xi_p=0$ with a radius $p<p_{\rm FB}$ for an infinite number of layers, $N\to \infty$. 
The effective mass is much larger than the characteristic band mass $m_3$ in 3D graphite. Indeed, we
have
$
m^*/m_3\sim \gamma_1/\gamma_4$
where we estimate $\hbar^2/(m_3 a_0^2) \sim \gamma_0$ as the conduction band width in graphite. We see that $m^*/m_3 \gg 1$. 
This dispersion is compared with the results of numerical diagonalization of $H({\bf K}, {\bf p})$ in Fig.~\ref{fig:2dnormalspectrum} using $\tilde \gamma_3 =0.066$, and $\tilde \gamma_4 =0.016$.

\paragraph{BdG equations} are constructed using the particle and hole Hamiltonians (\ref{H-particle}) and (\ref{H-hole}) coupled through the superconducting order-parameter field $\Delta$. As distinct from the quasiparticle energy measured from the chemical potential upwards, $E=\mu +\epsilon$, the energy of holes is measured from $\mu$ downwards, $E=\mu - \epsilon$. We have
\begin{eqnarray}
\sum_m \check \tau_3 \otimes \left[ \hat H_{nm}({\bf K},{\bf p}) -\mu  \delta_{nm}\right]\check \Psi_m + \check \Delta_n \check \Psi_n =\epsilon \check \Psi_n\ . \label{BdGHamilt}
\end{eqnarray} 
Here we introduce objects in the Nambu space
\[
\check \Delta_n
=\left(\begin{array}{lr} 0 & \Delta_n \\ \Delta^*_n &
0\end{array}\right)\ , \; \check \Psi_n =
\left(\begin{array}{c} \hat u_n 
\\ \hat v_n   \end{array}\right)\ , \; \check \tau_3 =\left(\begin{array}{lr} 1 & 0 \\ 0 &
-1 \end{array}\right) \ .
\]
Each component of the Nambu vector $\check \Psi_n$ 
is a pseudo-spinor. For analytical consideration we assume that $\Delta_n =0$ for $n\ne 1,N$. This is justified by the  numerical solution of the self-consistency equation using the full BdG equations \cite{KopninHeikkilaVolovik2011,KopninHeikkila12}.
In this case, Eq.~(\ref{BdGHamilt}) for $n\ne 1,N$ does not contain $\Delta$, so that one can use the
normal-state solution, Eq.~(\ref{psi}), where we have Nambu vectors $ \check A^\pm =
\left(A^\pm , \, B^\pm \right)^T$ instead of the corresponding scalars, and the Nambu matrix $\check \zeta$, Eq.~(\ref{zeta}), with $\check \tau_3 \epsilon$ instead of $\epsilon$.

At the outermost layers, the terms with $\hat u_0, \ \hat v_0$ and $\hat u_{N+1}, \ \hat v_{N+1}$  in Eq.~(\ref{BdGHamilt}) disappear.
The components which do not contain $\gamma_1$  yield
\begin{eqnarray}
\check \tau_3 \xi_{ p} \check A^-= (\tilde \epsilon - \check \tau_3 \tilde
\mu_p )\check A^+ - \check
\Delta_1 \check A^+ \ , \label{BdGeq-mu1}\\
\check \tau_3 \xi_{ p} \check A^+= (\tilde \epsilon - \check \tau_3 \tilde
\mu_p )\check A^- - \check \Delta_N \check A^- \ ,
\label{BdGeq-mu4}
\end{eqnarray}
where
$\tilde \epsilon =\epsilon \left(1-\tilde p^2\right)^{-1}$, 
$\tilde \mu_p =\mu_p \left(1-\tilde p^2\right)^{-1}$.
Equations (\ref{BdGeq-mu1}), (\ref{BdGeq-mu4}) provide the
surface-state spectrum 
and determine four independent surface states. 

If $\Delta_1 =\Delta_N$, the spectrum is
$
\tilde \epsilon^2 = \left(\tilde \mu_p \pm  \xi_p\right)^2+|\Delta|^2
$. If the number of layers $N$ is large, $\xi_p \to 0$ for $p<p_{\rm FB}$, the two surface states decouple
\begin{equation}
\tilde \epsilon _1^2 =\tilde \mu_p^2 +|\Delta|^2_1 \ , \; \tilde \epsilon _N^2 =\tilde \mu_p^2 +|\Delta|^2_N \ .\label{spectr-flat1}
\end{equation}
In this case, Eq.~(\ref{BdGeq-mu1}) at layer $n=1$ yields
$A^+ = U $, $B^+ = V $ or $A^+ = V $, $B^+ = U $ where
\begin{equation}
U =2^{-\frac{1}{2}}\left[ 1 + \tilde \mu_p/\tilde \epsilon\right]^{\frac{1}{2}}\ ,\;  V= 2^{-\frac{1}{2}}\left[ 1 -\tilde \mu_p/\tilde \epsilon \right]^{\frac{1}{2}}\ . \label{UV}
\end{equation}

\paragraph{Surface superconductivity.}

The surface states discussed above form a basis for the superconducting gap localized near outer surfaces.
Within the mean-field approximation, the gap at layer $n$ is determined by the self-consistency equation.
As was shown in Ref.~\onlinecite{KopninHeikkilaVolovik2011}, the surface states dominate due to a much larger DOS. For a large number of layers when $\xi_p=0$, the self-consistency equation for the gap at the surface takes the form
\begin{eqnarray}
1&=&\frac{W}{d} \int_{\rm FB}\frac{d^2 p}{(2\pi \hbar)^2} 
\frac{(1- \tilde p^2)}{\tilde \epsilon} \tanh\frac{\epsilon}{2T}\ .\quad \label{Delta-gen} 
\end{eqnarray}
Here we used Eq.~(\ref{UV}) to find the integrand, $\epsilon$ is one of the spectral branches in Eq.~(\ref{spectr-flat1}), the integration is carried out over momenta within the flat band, $p<p_{\rm FB}$, and $W$ is the 3D coupling potential. This is the central result of our Letter. The superconducting coupling is described by the energy 
$
g= (W/d)p_{FB}^2/ \hbar^2.
$
It can also be expressed in terms of the usual BCS coupling constant $\lambda =\nu_3 W$ where $\nu_3 =m_3 p_{3F}/2\pi^3 \hbar^3$ is the 3D density of states and $p_{3F}$ is the Fermi momentum in 3D graphite. Assuming the conduction band width in 3D graphite of the order of $\gamma_0$ we have $g/\gamma_1 \sim \lambda (\gamma_1 /\gamma_0)$ if $ \hbar /a_0p_{3F} \sim 1$. 

\begin{figure}[t]
\centering
\includegraphics[width=0.8\linewidth]{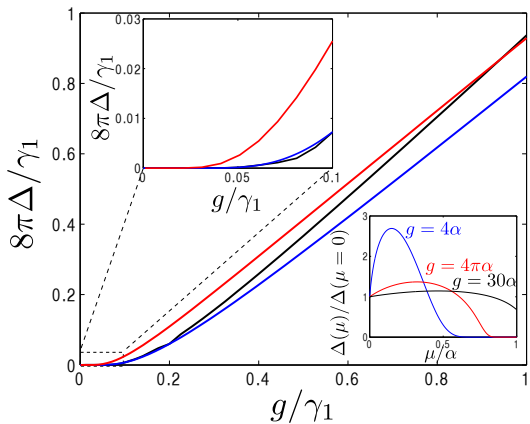}
\caption{Self-consistent surface gap vs. coupling $g$. The black line (in the middle) shows the results based on the exact diagonalization of the BdG equations for $N=20$ with $\mu=0$, the blue (bottom) line is Eq.~\eqref{Delta-gen} at $\mu=0$ and the red (top) line corresponds to $\mu=\mu_{\rm opt}$ that maximizes $\Delta$ for given $g$. For large $g$, the gap tends towards the flat-band limit $\Delta \propto g$. For $g \lesssim 4\pi \alpha$ (upper inset), the gap is exponentially suppressed, $\Delta \propto \exp(-4\pi\alpha/g)$. The lower inset shows the (normalized) gap as a function of $\mu$ for a few values of $g$. }
\label{fig:selfconsdelta}
\end{figure}

The overall behavior of $\Delta$ vs.\ the coupling energy is plotted in Fig.~\ref{fig:selfconsdelta}. Let us consider first the resulting $\Delta$ for zero doping $\mu=0$. The quadratic dispersion due to $\gamma_4$ comes with an energy scale $\alpha=2\tilde \gamma_4 \gamma_1$, which determines a crossover between exponentially suppressed and flat band superconductivity. For $g \gg 4\pi \alpha$ and for zero doping, Eq.~\eqref{Delta-gen} yields the flat-band result \cite{KopninHeikkilaVolovik2011}, $\Delta =g/8\pi$ for $T=0$, and the critical temperature satisfying $\Delta=3k_B T_C$. 
Due to its linear dependence on the interaction strength, {\it the critical temperature is proportional to the area of the flat band and can be essentially
 higher than that in the bulk}. Doping in the flat band regime destroys the surface superconductivity \cite{KopninHeikkilaVolovik2011}. Both $\Delta_0$ and $T_c$ vanish at the critical doping level  $|\mu| =2k_B T_c$.

For $g \ll 4\pi \alpha$ the weak dispersion Eq.~(\ref{eq:normaldispersion}) with a heavy mass $m^*$ dominates. The integral in Eq.\ (\ref{Delta-gen}) is logarithmic which results in a BCS-like expression (for $T=0$)
\[
\Delta =[\alpha^2/(\alpha -\mu)] e^{-1/\lambda_2}  \ , \; \lambda_2= g(1-\mu/\alpha)^2/4\pi \alpha 
\]
where 
$
\alpha=2\gamma_1\tilde \gamma_4 
$. 
The estimate for $g$ gives
$
\lambda_2 \sim \lambda (\gamma_1/\gamma_4)$. {\it This is a much larger coupling constant than $\lambda$ for bulk superconductivity.} The gap disappears at $\mu =\alpha$. The crossover from the BCS-like to the flat-band regime occurs at  $g\sim 4\pi\alpha$,
and $\Delta \sim \alpha$.
The coherence length
$
\xi_0  =\hbar v_g/\Delta  \sim a_0(\gamma_0/\gamma_1) e^{1/\lambda_2}
$ is much longer than the interatomic distance $a_0$.  These analytical results are compared  in Fig.~\ref{fig:selfconsdelta} to the numerical solution of the self-consistency equation using the full Hamiltonian, Eq.~(\ref{BdGHamilt}). 
In contrast to the flat-band and the BCS regimes, the gap in the intermediate region $\alpha \sim g$ is enhanced by an optimum doping, i.e., the critical temperature is very sensitive to the presence of impurities (lower inset in Fig.~\ref{fig:selfconsdelta}). This complies with the reports of high-temperature superconductivity in doped graphite, Ref.~\onlinecite{Kopelevich01}.

\paragraph{Effect of fluctuations.}

The quality of the mean-field approximation used above is determined by the Ginzburg number Gi which is a measure of the relative magnitude of order-parameter fluctuations. For usual 3D superconductors ${\rm Gi}\ll 1$ due to a small ratio of the critical temperature to characteristic energy of electrons (i.e., the Fermi energy). Here we demonstrate that the mean field approach also works well when the quadratic dispersion dominates over the flat band. In this case the fluctuation free energy density for $T$ not too close to $T_c$ is
$F_1 \sim \nu_2 \Delta_1^2/2$,
where $\nu_2 =m^*/2\pi \hbar^2$ is the 2D DOS and the effective mass $m^*$ is determined by Eq.~(\ref{mass}).
The energy of an area $\pi \xi_0^2$ with a radius of $\xi_0 =\hbar v_g \Delta_0^{-1}$ is
$
{\cal F}_1\sim \pi \xi_0^2 F_1 = \tilde \gamma_4\gamma_1(\Delta_1^2/\Delta_0^2)
$,  
where $\Delta_0$ is the mean-field gap. Since $ {\cal F}_1\sim T$ we find
$
\Delta_1^2/\Delta_0^2={\rm Gi} \sim  T_c/ \tilde \gamma_4\gamma_1 $.
When the quadratic dispersion dominates, one has $T_c \ll \gamma_1\tilde \gamma_4 $ with the Ginzburg number ${\rm Gi}= e^{-1/\lambda_2}\ll 1$, thus the average fluctuation of the order parameter is small compared to its mean-field value. However, at the crossover to the flat band regime, the fluctuation becomes of the same order as the mean-field $\Delta_0 \sim \gamma_1\tilde \gamma_4$. Therefore, the mean-field approach is not exact for the flat-band regime. Nevertheless, it is used here as an initial step towards a full theory of high temperature surface superconductivity.

\paragraph{Summary.}

Rhombohedral graphite is a promising candidate for high-temperature surface superconductivity due to its (approximate) topologically protected flat band. Besides surfaces, similar type of superconductivity may arise around stacking faults and interfaces between differently stacked regions of graphite, as long as the system contains more than a few layers of RHG regions \cite{KopninHeikkila12}. Recent observations of high-temperature superconductivity in graphite \cite{Kopelevich01,Esquinazi08,Ballestar12,Esquinazi12} are compatible with surface or interface superconductivity described by our theory if there are RHG regions embedded inside otherwise Bernally stacked graphite. Our predictions can be used for search or for an artificial fabrication of layered and/or twinned systems with high- and even room-temperature superconductivity. With the hopping parameters used above, the crossover between the flat-band and BCS-like regimes takes place around $g_c \sim 4\pi \alpha \approx 0.39 \gamma_1 \approx 0.15$ eV (see Fig.~\ref{fig:selfconsdelta} and Appendix) corresponding to the mean-field $T_c(g_c) \approx 20$ K. For $g>g_c$ we thus find $T_c \sim (g/\gamma_1) \times 50$ K. This is much greater than the expected gap in the bulk for the same magnitude of coupling.

\acknowledgements

We thank G.\ Volovik for helpful comments and the collaboration that initiated this project. We also acknowledge fruitful discussions with F. Mauri.
This work is supported in part by the Academy of Finland through its
Centre of Excellence Program (projects no. 250280 and 251748 and by the European Research Council (Grant
No. 240362-Heattronics).


\section{Appendix: Computational details}

The density-functional theory calculations on rhombohedral graphene slabs were performed using the all-electron FHI-aims code \cite{AIMS}. We used the Perdew-Burke-Ernzerhof (PBE) exchange-correlation functional for all calculations, and took into account the van der Waals interaction using the approach by Tkatchenko and Scheer \cite{Tkatchenko-Scheer}. The Brillouin zone was sampled using a 48$\times$48$\times$1 $k$-point grid. "Tight" basis defaults as defined in the FHI-aims distribution were used for the all-electron description of the carbon atoms.

The in-plane lattice parameter was optimized for monolayer graphene and it was found to be 2.466~\AA{}.  The same lattice parameter was used for the rhombohedral slabs but the distance between adjacent layers was allowed to relax freely until forces acting on atoms were less than 0.001~eV/\AA{}. This resulted in variations of the order 0.01~\AA{} in the interlayer distances, the outer layers being slightly expanded from the optimized interlayer distance 3.332~\AA{} of rhombohedral graphite bulk. The use of the lattice parameter of graphene leads to negligible errors, as the difference to the optimized bulk parameter, 2.462~\AA{}, is very small, and the multilayer structures are expected to interpolate between the bulk and monolayer limits.

\section{Fitting the tight-binding model}

\begin{figure*}[t]
\includegraphics[width=0.85\columnwidth]{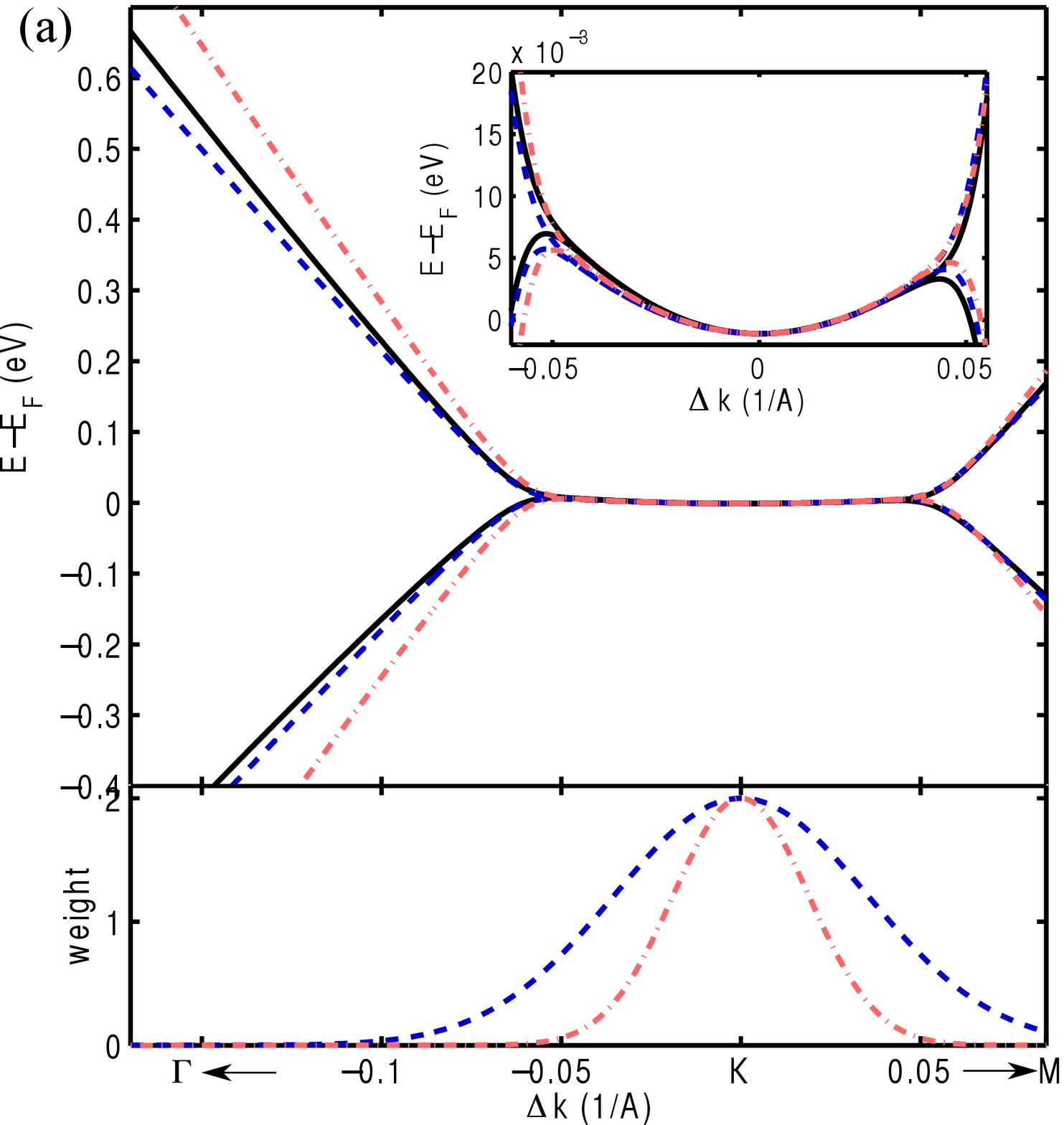}
\includegraphics[width=0.85\columnwidth]{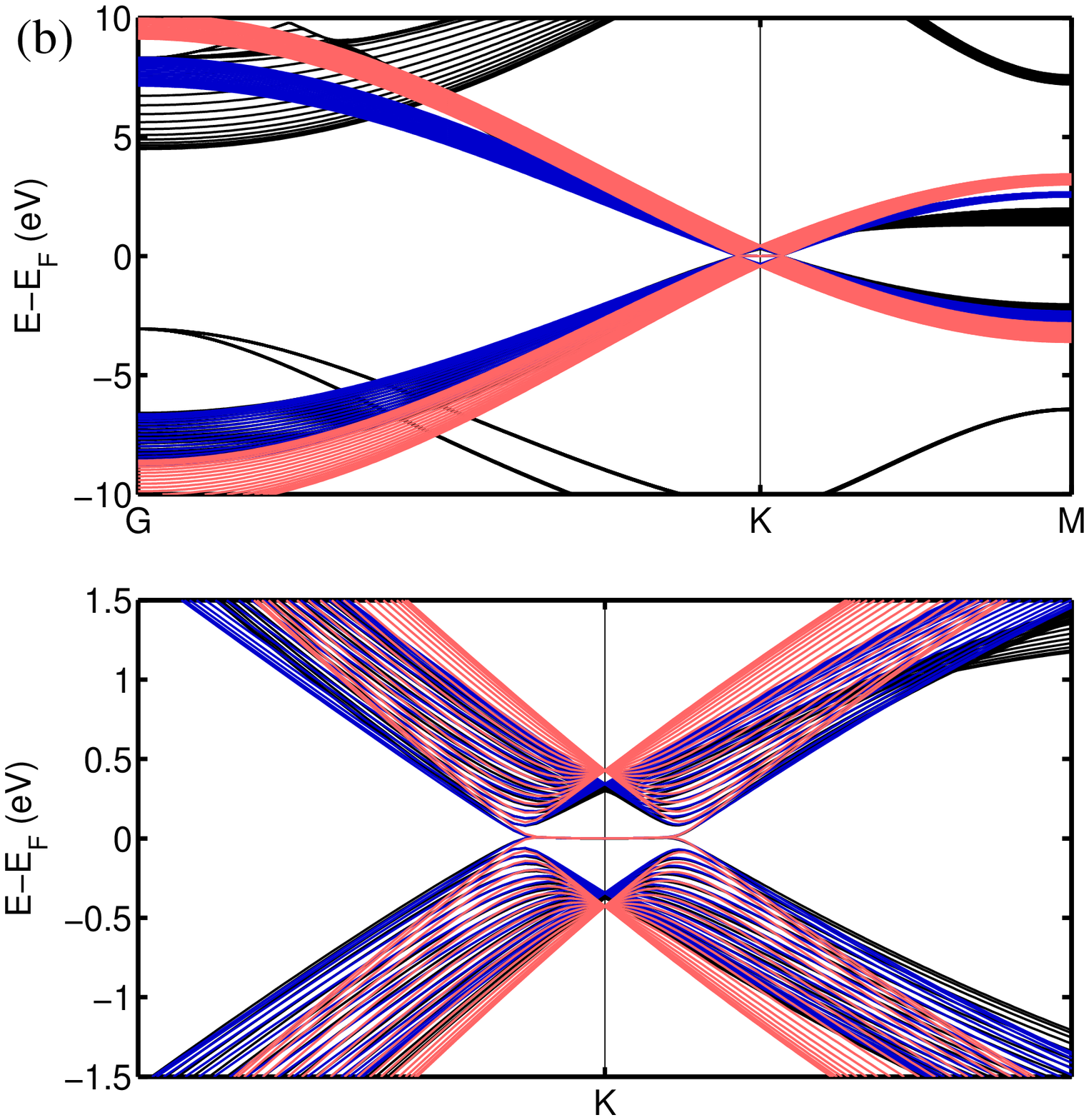}


\caption{\label{fig:suppl1} Fitting of the tight-binding parameters for a 20-layer ABC-stacked graphene slab. (a) Comparison of the bands close to the $K$-point, as well as the form of the weighing function used in the least squares fit. Inset shows a close-up of the parabolic region of the flat band.  (b) A comparison of the tight-binding and DFT band structures along the $\Gamma-K-M$ lines for the two found optima. Black -- DFT, blue -- tight-binding fit 1, red -- tight-binding fit 2. }
\end{figure*}

The band structures were calculated along the $\Gamma-K-M$ direction in the Brillouin zone, sampling the lines between (0.3,0.3) and (1/3,1/3), as well as between (1/3 and 1/3) and (0.3,0.35) using 200+200 $k$-points. The band structure in the vicinity of the $K$-point was fitted to a tight-binding model, in which hoppings between nearest neighbors in-plane ($\gamma_0$) and out-of-plane ($\gamma_1$), as well as between next-nearest neighbors out-of-plane ($\gamma_3$ and $\gamma_4$) were included. The fit was concentrated to the region around $K$ using a Gaussian function to weigh the squared error between the tight-binding and DFT bands, and the minimal energy of the parabolic region was shifted to match the corresponding DFT energy. The width of the Gaussian was chosen such that the weight outside the flat band region was practically zero. Fig.~\ref{fig:suppl1}(a) demonstrates a comparison between the DFT band structure and the corresponding tight-binding fit, as well as the shape of the weighing function on a 20-layer slab. 

Two parameter sets with comparable agreement around the $K$-point were found when the width of the weighing Gaussian was altered. As Fig.~\ref{fig:suppl1}(b) and (c) show, one of them better captures the overall band structure. It was thus chosen for the further calculations. Both parameter sets are reported in Table~\ref{table:suppl1}. The differences in the parameters is insignificant for the considerations of the surface superconductivity in the main paper, and they both give reasonable Fermi velocities ($v_{F,1}$ =0.84$\cdot 10^6$~m/s and $v_{F,2}$ =1.04$\cdot 10^6$~m/s, respectively) . 

\begin{figure}[hb]
\includegraphics[width= 0.84\columnwidth]{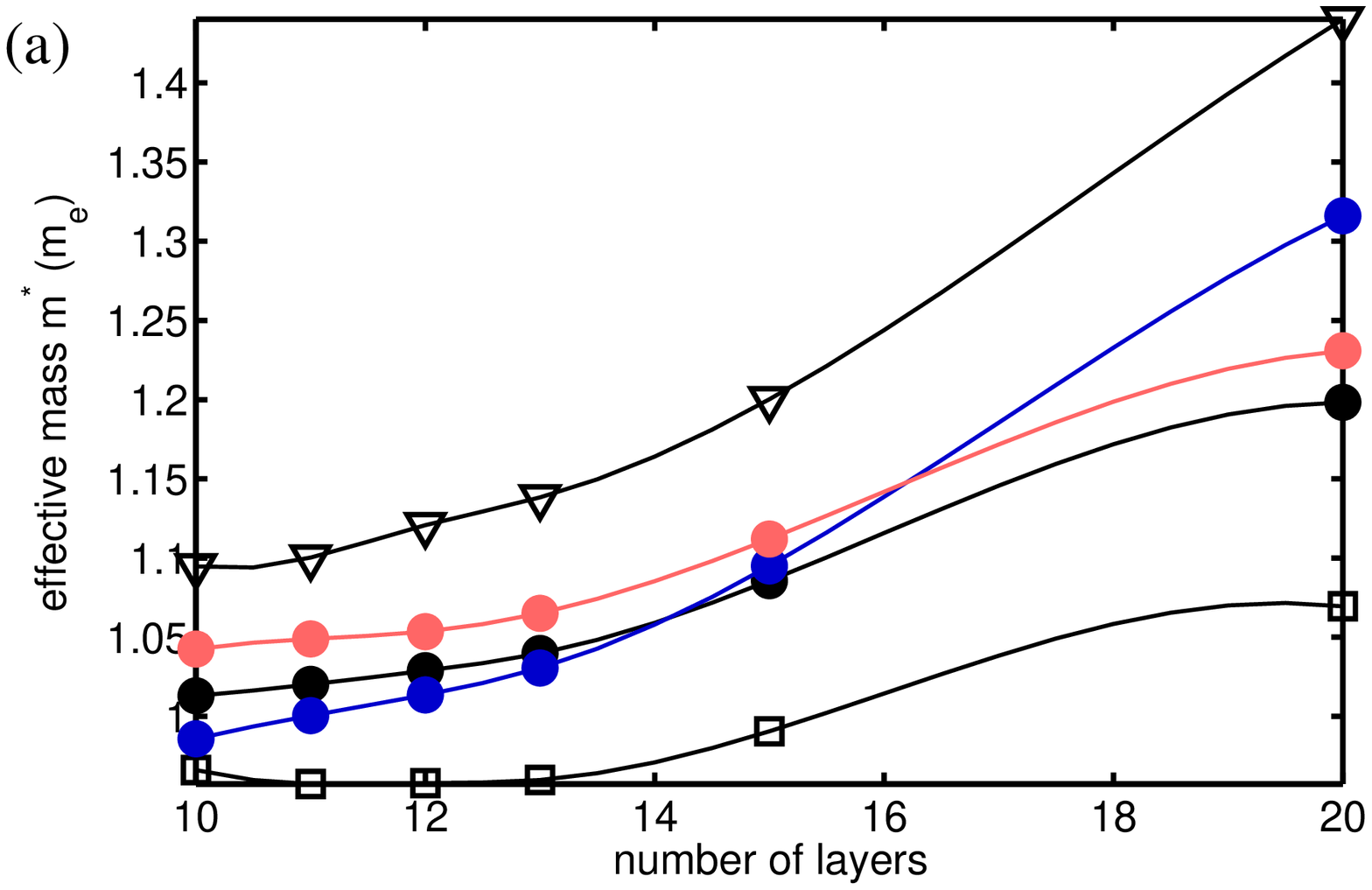}
\includegraphics[width = 0.84\columnwidth]{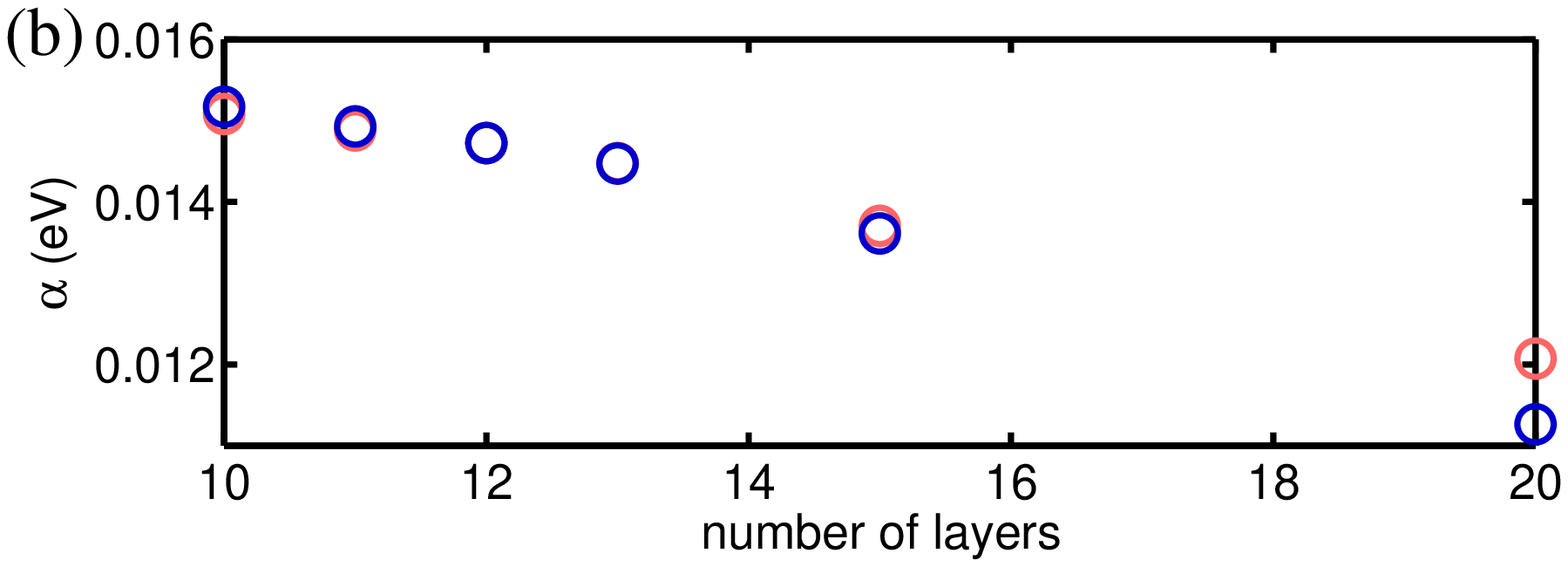}
\caption{\label{fig:suppl3} (a) Effective mass $m^*$ obtained from the parabolic fit to the DFT bands (black), and calculated from the fitted tight-binding parameters using  [Eq.~(8)]. The black open triangles refer to $m^*$ from DFT correponding to the $K-\Gamma$ direction, and open squares to $K-M$ direction. Filled black circles result from a fit using both directions. The lines are a guide to the eye. (b) Tight-binding fits for the parameter $\alpha$. In both plots, blue symbols refer to fit 1 and red symbols to fit 2 (Table \ref{table:suppl1}).}  \end{figure}

\begin{table}
\caption{\label{table:suppl1} Tight-binding parameters as a function of the slab thickness. Both sets show comparable agreement around $K$-point in the parabolic regime but fit 1 better captures the overall shape of the few-layer ABC-stacked graphene slabs.}
\begin{tabular}{c|c|c}
 & fit 1 & fit2 \\
 layers & $\gamma_0$, $\gamma_1$, $\gamma_3$, $\gamma_4$ &$\gamma_0$, $\gamma_1$, $\gamma_3$, $\gamma_4$\\
\hline
\hline
10&2.62, 0.35, 0.17, 0.06&3.21, 0.44, 0.16, 0.06\\
15&2.59, 0.35, 0.17, 0.05& 3.21, 0.43, 0.15, 0.05\\
20&2.58, 0.34, 0.17, 0.04&3.21, 0.43,  0.12, 0.05 \\
\end{tabular}
\end{table}

\section{Determining the effective mass directly from  a parabolic fit to the DFT bands around $K$}

Additionally, the DFT-calculated dispersion very close to the K-point was fitted directly using a parable [Eq.~(8) in the main text], and a value for the effective mass $m^*$ and chemical potential $\mu$ were extracted separately for both the $K-\Gamma$ and $K-M$ directions, in addition to a mean value fitted simultaneously to both directions. These values were compared to those obtained from the tight-binding fit. Fig.~\ref{fig:suppl3}(a) shows the $m^*$ and $\mu$ as a function of the slab thickness as well as those calculated from Eq.~(9) based on the fitted tight-binding parameters $\gamma_i$.  The effective mass increases with an increasing number of layers, as the breadth of the flat band increases.

In DFT, there is also a small linear component in the parabolic fit $E-E_F = \beta_0 (\Delta k)^2+\beta_1 (\Delta k)-\mu$ due to the longer-range couplings neglected in the tight-binding model. The effect of this component is more important further away from K-point, as it decreases when the fitting region is made narrower.  In DFT, the minimum of the parabolic bands does not lie exactly at the Fermi energy. This finite doping is, however, small, $\mu/\alpha < 0.1$ in all DFT calculations.

The parameter $\alpha$ determines the crossover between the  BCS- and flat-band regimes. It is worth noting that even though the values for the hopping amplitudes differ in the tight-binding parameters sets, both yield comparable values for $\alpha$, as illustrated in Fig.~\ref{fig:suppl3}(b). Moreover, the parameter set given in Ref.~\onlinecite{reviewCastroNeto09b}, $\gamma_0$ = 3.2~eV, $\gamma_1$ = 0.39~eV, $\gamma_3$ = 0.315~ eV and $\gamma_4$ = 0.044~eV, gives $\alpha = 0.012$~eV, showing that this value is quite universal.

\end{document}